\PassOptionsToPackage{nonamebreak}{natbib} 
\documentclass[a4paper,fleqn,usenatbib]{mnras}

\usepackage{newtxtext,newtxmath}
\usepackage[T1]{fontenc}
\usepackage{ae,aecompl}

\usepackage{float}
\usepackage{array}
\usepackage{relsize}
\usepackage{graphicx}	% Including figure files
\usepackage{amsmath}	% Advanced maths commands
\usepackage{amssymb}	% Extra maths symbols
\usepackage{lscape}

\def\aap{A\&A}%

\def\aj{AJ}%
\def\apj{ApJ}%
\def\apjl{ApJ}%
\def\apjs{ApJS}%
\def\ao{Appl.~Opt.}%
\def\jcap{J. Cosmology Astropart. Phys.}%
\def\mnras{MNRAS}%
\def\nar{New A Rev.}%
\def\nat{Nature}%
\def\pasp{PASP}%
\def\prd{Phys.~Rev.~D}%
\def\procspie{Proc.~SPIE}%
\title[Dark halo structure in Carina dSph]{Dark halo structure in the Carina dwarf spheroidal galaxy:
joint analysis of multiple stellar components}

\author[K. Hayashi et al.]{Kohei Hayashi$^{1,2}$\thanks{Contact e-mail: kohei.hayashi@ipmu.jp},
Michele Fabrizio$^{3,4}$, Ewa L. {\L}okas$^{5}$, Giuseppe Bono$^{6,3}$, \newauthor
Matteo Monelli$^{7,8}$, Massimo Dall'Ora$^{9}$ and Peter B. Stetson$^{10}$
\\
$^{1}$National Astronomical Observatory of Japan, 2-21-1 Osawa, Mitaka, Tokyo 181-8588, Japan\\
$^{2}$Institute for Cosmic Ray Research (ICRR), The University of Tokyo, Chiba 277-8583, Japan\\
$^{3}$INAF - Osservatorio Astronomico di Roma, Via di Frascati 33, 00078 Monte Porzio Catone (Roma), Italy\\
$^{4}$Space Science Data Center - ASI, Via del Politecnico SNC, 00133 Roma, Italy\\
$^{5}$Nicolaus Copernicus Astronomical Center, Polish Academy of Sciences, Bartycka 18, 00-716 Warsaw, Poland\\
$^{6}$Dipartimento di Fisica, Universita di Roma ``Tor Vergata'', Via della Ricerca Scientifica 1, I-00133 Roma, Italy\\
$^{7}$Instituto de Astrof\'{i}sica de Canarias, Calle Via Lactea, E-38205, La Laguna, Tenerife, Spain\\
$^{8}$Universidad de La Laguna, Dpto. Astrof\'{i}sica, E-38206, La Laguna, Tenerife, Spain\\
$^{9}$INAF-Osservatorio Astronomico di Capodimonte, Salita Moiariello 16, I-80131 Napoli, Italy\\
$^{10}$Dominion Astrophysical Observatory, NRC-Herzberg, National Research Council, 5071 West Saanich Road, Victoria, BC V9E 2E7, Canada
}
\date{Accepted XXX. Received YYY; in original form ZZZ}

\pubyear{2018}

\begin{document}
\label{firstpage}
\pagerange{\pageref{firstpage}--\pageref{lastpage}}
\maketitle

\begin{abstract}
Photometric and spectroscopic observations of the Carina dSph revealed that this galaxy contains two dominant stellar 
populations of different age and kinematics. The co-existence of multiple populations
provides new constraints on the dark halo structure of the galaxy, because different populations should be in
equilibrium in the same dark matter potential well. We develop non-spherical dynamical models including such multiple
stellar components and attempt to constrain the properties of the non-spherical dark halo of Carina. We find that
Carina probably has a larger and denser dark halo than found in previous works and a less cuspy inner dark matter
density profile, even though the uncertainties of dark halo parameters are still large due to
small volume of data sample.
Using our fitting results, we evaluate astrophysical factors for dark matter annihilation and decay and find that
Carina should be one of the most promising detectable targets among classical dSph galaxies. We also calculate stellar
velocity anisotropy profiles for both stellar populations and find that they are both radially anisotropic in the inner
regions, while in the outer regions the older population becomes more tangentially biased than the intermediate one.
This is consistent with the anisotropy predicted from tidal effects on the dynamical structure of a satellite galaxy
and thereby can be considered as kinematic evidence for the tidal evolution of Carina.
\end{abstract}

\begin{keywords}
galaxies: Carina -- galaxies: dwarf -- galaxies: kinematics and dynamics -- dark matter
\end{keywords}

%%%%%%%%%%%%%%%%%%%%%%%%%%%%%%%%%%%%%%%%%%%%%%%%%%
%%%%%%%%%%%%%%%%% BODY OF PAPER %%%%%%%%%%%%%%%%%%
%%% Sec.1 %%%%%%%%%%%%%%%%%%%%%%%%%%%%%%%%%%%%%%%%
%%%%%%%%%%%%%%%%%%%%%%%%%%%%%%%%%%%%%%%%%%%%%%%%%%
\section{Introduction} \label{sec:intro}

Dwarf spheroidal (dSph) galaxies in the Local Group are an excellent test-bed for understanding the fundamental
properties of dark matter and galaxy formation processes involving this non-baryonic matter. In the area of
particle physics, these satellites have drawn special attention as ideal sites for obtaining limits on
particle candidates of TeV-scaled dark matter~\citep[e.g.,][]{Lake1990,Wal2013,Geretal2015,Hayetal2016}. This is
because these galaxies have high dynamical mass-to-light ratios ($M/L\sim10-1000$), that is, are the most dark matter
dominated systems~\citep[][and the references therein]{McC2012}. Moreover, these galaxies are sufficiently close to
measure line-of-sight velocities for their resolved member stars by high-resolution spectroscopy, and this
kinematic information enables us to study structural properties of dark matter haloes in less massive galaxies in
detail.

Using these high-quality data, the studies of dSphs have suggested some intriguing properties of dark halos in dSphs;
some of them may have cored or at least shallower cuspy dark matter
profiles~\citep[e.g.,][]{Giletal2007,Batetal2008,WP2011}, are most likely to have non-spherical dark
halo~\citep[][]{HC2012,HC2015b}, and their dark
halos show some universal properties~\citep[][]{SB2000,Stretal2008,Boyetal2010,Saletal2012,HC2015a,KF2016,Hayetal2017}. In particular,
cored dark halo profiles in dSphs are in disagreement with those predicted from dark matter simulations based on
cold dark matter~(CDM) models, which showed that the density profile of a dark halo of any mass is well fit by a
Navarro-Frenk-White profile~\citep[hereafter ``NFW'';][]{NFW1996,NFW1997}. This issue is the so-called ``core-cusp''
problem, and it is yet a matter of ongoing debate. In order to solve or alleviate the above issue, a possible solution
has been proposed that relies on a transformation mechanism from cusped to cored central density. Recent
high-resolution cosmological $N$-body and hydrodynamical simulations in the context of CDM models have shown that inner
dark halo profiles at dwarf-galaxy mass scale could be transformed to cored ones due to the effects of energy feedback
from star-formation activity such as radiation energy from massive stars, stellar winds and supernova
explosions~\citep[e.g.,][]{Govetal2012,Madetal2014,DiCetal2014a,DiCetal2014b,OM2014,Onoetal2015}.

Another possible solution is to replace CDM with alternative dark matter models such as warm dark matter or
self-interacting dark matter models, which can form cored and less dense central dark matter profiles on less massive
scales without baryonic effects~\citep[e.g.,][]{Andetal2013,Bozetal2016,SS2000,Vogetal2012,Elbetal2015}. However,
because of the dependence on the observational and theoretical uncertainties, it is difficult to discriminate
between these two different transformation mechanisms on the basis of observational facts, hence the core-cusp problem
still persists.

On the other hand, current kinematic studies of dSphs are unable to determine precisely dark
halo structures in these galaxies because of the presence of degeneracy in mass models, which stems from the fact that
only projected kinematic information is available and dynamical models are incomplete. For example, kinematic studies
typically treat dSph galaxies as spherically symmetric systems with constant velocity anisotropy along the radii.
However, in such models, there is a degeneracy between the velocity anisotropy of stars and dark matter density
profiles~\citep[e.g.,][]{Evaetal2009,Waletal2009c}. Even for axisymmetric mass models, similar degeneracy between
stellar velocity anisotropy and the global shape of dark halo exists~\citep[][]{Cap2008,HC2015b,Hayetal2016}. The
studies of dark matter in dSph galaxies have been hampered by these degeneracies. In order to disentangle this
ambiguity, at least in part, so-far unknown kinematical information on dSphs and/or more general dynamical models are
required.
Recently, owing to the outstanding quality of {\it Gaia}~\citep[e.g.,][]{Gaia1,GaiaDR2a,GaiaDR2b}, the precision proper motions of nearby bright stars are available.
Using the proper motions of 15 member stars in Sculptor dSph based on data from the {\it Gaia} and the {\it Hubble Space Telescope}, \citet{Masetal2018} measured the internal motions of the stars and obtained the limit on the velocity anisotropy. Although there exists a large uncertainty on this anisotropy with the paucity of data sample, a number of proper motion data~from the {\it Gaia} and future astrometry observations~(e.g. Wide Field Infrared Survey Telescope~\citep{WFIRST2015}, NFIRAOS at Thirty Meter Telescope~\citep{TMT2014} and MICADO/MAORY at Extremely Large Telescope~\citep{EELT2017} and so on) will enable us to tackle the issues of the degeneracies head-on.

Recent spectroscopic observations found that some classical dSphs (Sculptor, Fornax and Sextans) exhibit multiple
chemo-dynamical sub-populations~\citep[][]{Toletal2004,Batetal2006,Batetal2008,Batetal2011}. For instance,
\citet{Batetal2008} found that in the Sculptor dSph metal-rich stars are centrally concentrated and have colder
kinematics with the line-of-sight velocity dispersion decreasing with radius, whereas metal-poor ones
are more extended and have hotter kinematics and an almost constant dispersion profile. From the perspective of
dynamical modeling, the presence of these intriguing chemo-dynamical stellar properties provides new constraints on the
dark halo structure of dSphs. This is because these kinematically different populations should settle in the same dark
matter potential well, and thus the co-existence of multiple populations enhances our ability to infer the inner
structure of the dark matter halo~\citep[][]{Batetal2008,WP2011,AmE2012,AgE2012,Amoetal2013,BH2014,Stretal2017}.

\citet{Batetal2008}~were the first to attempt to set constraints on the inner slope of dark matter profile in the
Sculptor dSph using multiple stellar components. Using spherical Jeans models, they found that cored dark halo models
provide a better fit than NEW models, which however could not be ruled out. Later, \citet{WP2011}~have separated
multiple stellar components using a combined likelihood function for spatial, metallicity and velocity distributions,
and then constrained the dark matter slopes of the Sculptor and Fornax dSphs through the mass slope using metal-rich
and metal-poor populations and assuming a spherical stellar system. They concluded that both Fornax and especially
Sculptor have central cores, ruling out the NFW profile at high statistical significance.
However, if a dSph is not spherical, assuming sphericity in their method may lead to a strong bias and the inferred slope turns
out to depend on the line-of-sight~\citep{Kowetal2013,Genetal2018}. In contrast, adopting a separable model for the
distribution function of an equilibrium spherical system \citet{Stretal2017}~showed that the two chemo-dynamical
subpopulations in Sculptor are consistent with an NFW profile. Furthermore, \citet{BH2014}~applied Schwarzschild
methods to two stellar subpopulations in Sculptor and concluded that the fitting results with the NFW model are
indistinguishable from those with a cored one. Therefore, the remarkable debate as to whether the central
regions of halos in dSphs are cored or cusped is still ongoing.

All these studies assumed, however, that the stellar and dark components are spherical, even though we know both
from observations and theoretical predictions that these components should be actually non-spherical.
\citet{Zhuetal2016}~applied the discrete Jeans anisotropic multiple Gaussian expansion model~\citep[][]{Watetal2013} to
the chemo-dynamical data of the Sculptor dSph assuming that stellar distributions are free to be axisymmetric but a
dark matter halo is spherical. On the other hand, \citet{HC2015b} have constructed axisymmetric mass
models based on axisymmetric Jeans equations, including stellar velocity anisotropy and applied these models to
line-of-sight velocity dispersion profiles of the luminous dSphs associated with the Milky Way and Andromeda.
\citet[][hereafter H16]{Hayetal2016}~applied the generalized axisymmetric mass models developed by~\citet{HC2015b} to
the most recent kinematic data for the ultra faint dwarf galaxies as well as classical dSphs, but they performed an
unbinned analysis for the comparison between data and models, unlike~\citet{HC2015b}. However, these non-spherical mass
models have not yet been applied to two-component systems in any dSphs.

In this paper, following the axisymmetric mass models developed by~H16 and using an unbinned analysis, we develop the
multiple stellar component mass models to obtain stronger limits on dark halo structures in dSphs and apply these models
to the line-of-sight velocity data of two different age populations in the Carina dSph. As a result of a long-term
photometric and spectroscopic observation project for Carina~\citep[Carina
Project:][]{Daletal2003,Monetal2003,Fabetal2011,Fabetal2012,Copetal2013,Monetal2014,Fabetal2015,Fabetal2016}, it has
been shown that the stars in this galaxy can be separated into an intermediate-age and an old-age population
using the color-magnitude diagram derived from these photometric data. The spectroscopic data for a fraction of
these stars were then obtained using Very Large Telescope~(as described in next section in more detail). Therefore,
in this work we focus on the constraints on the dark halo structure in Carina that can be obtained using the
mass models we developed and recent photometric and spectroscopic data for the two stellar populations. Moreover, the
important point in this work is that we attempt for the first time to set constraints on a non-spherical dark halo of
Carina using the data for its two stellar sub-populations.

\begin{figure*}
	\begin{center}
	\includegraphics[scale=0.18]{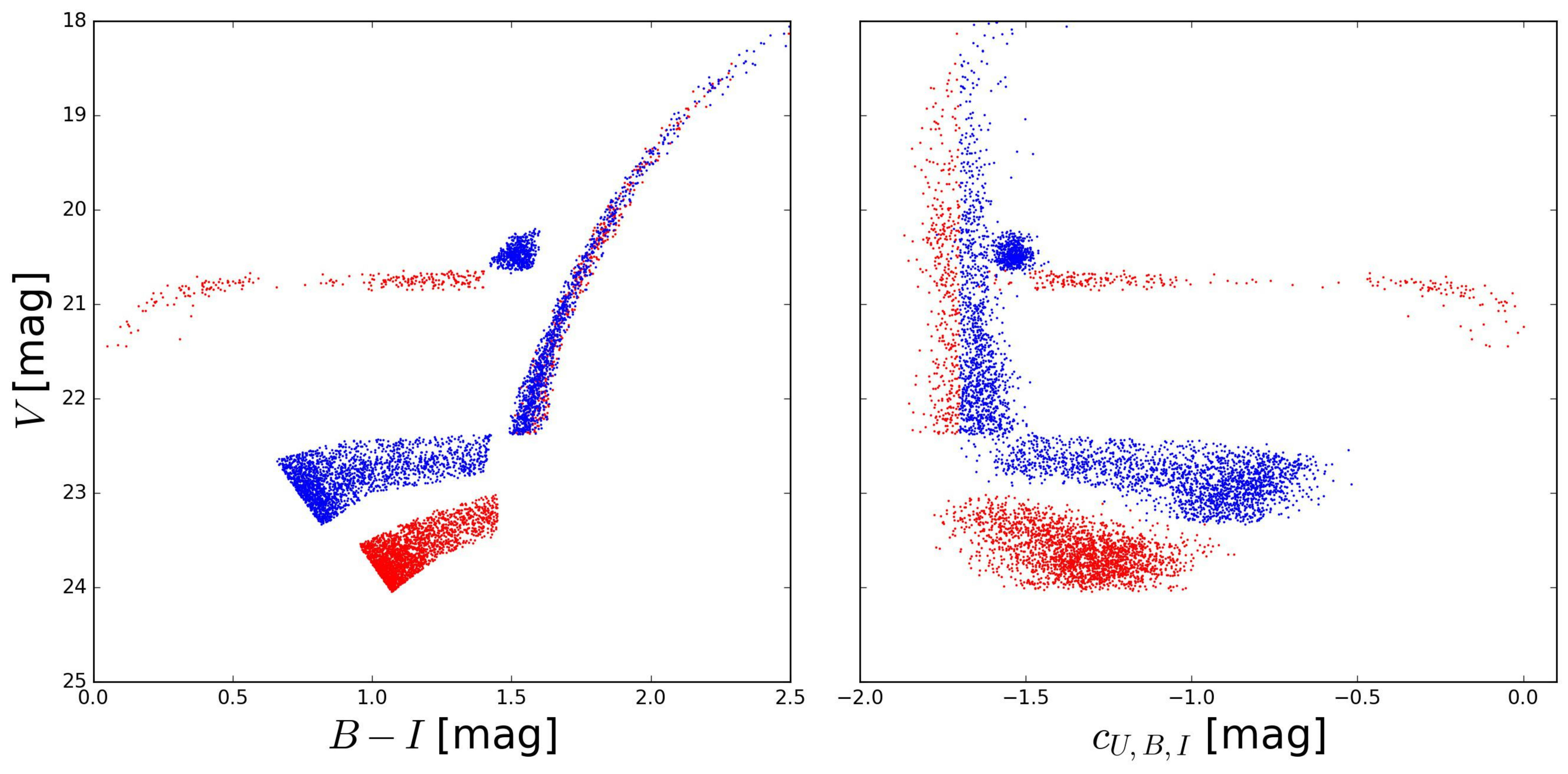}
	\end{center}
\caption{Left -- Photometric selection of intermediate-age (blue) and old (red) candidate Carina stars based on the
$V$~versus~$(B-I)$ CMD. 
The former includes the red clump stars and bright sub-giant branch stars, while the latter one includes horizontal
branch stars and faint sub-giant branch stars. Right -- Same as the left panel, but in the $V$~versus~$c_{U,B,I}$
diagram. The intermediate and old-age candidates selected in the left panel are also plotted with blue and red
symbols.}
\label{fig:CMD}
\end{figure*}

The paper is organized as follows.
In Section~2, we describe the photometric and spectroscopic data for Carina dSph.
In Section~3, we explain axisymmetric models for density profiles of stellar and dark halo components based on an
axisymmetric Jeans analysis.
In Section~4, we introduce the method of fitting the data and the joint likelihood function we adopted.
In Section~5, we present the results of the fitting and compare these with the ones from the single likelihood
function analysis.
In Section~6, we discuss astrophysics factors for dark matter annihilation and decay and the stellar velocity
anisotropy profile for each population.
Finally, conclusions are presented in Section~7.

%%% Table 1 %%%
\begin{table}
	\centering
	\caption{Parameter constraints for stellar surface density distributions of the two populations in Carina. The ``ALL'' column indicates fitting results using all~(INTERMEDIATE$+$OLD) stars in Carina.}
	\label{table1}
    \scalebox{0.8}[0.8]{
	\begin{tabular}{lrrr} % four columns, alignment for each
		\hline\hline
		                      & INTERMEDIATE & OLD & ALL\\
		\hline
		{\bf S\'ersic profile} & & &\\    
		  Ellipticity         & $0.29^{+0.02}_{-0.03}$      & $0.32^{+0.04}_{-0.04}$     & $0.33^{+0.02}_{-0.02}$    \\
		  $R_s$      [arcmin] & $11.7^{+0.7}_{-0.7}$        & $14.3^{+1.4}_{-1.2}$       & $12.5^{+0.9}_{-0.8}$      \\
		  $R_s$      [pc]     & $357.4^{+21.4}_{-21.4}$     & $436.8^{+36.7}_{-42.8}$    & $381.8^{+27.5}_{-24.4}$   \\
		  $m$                 & $0.50^{+0.04}$              & $0.52^{+0.09}_{-0.02}$     & $0.53^{+0.05}_{-0.03}$    \\
		  reduced-$\chi^2$    & $1.16$                      & $1.10$                     & $1.04$                    \\
		{\bf Plummer profile} & & &\\    
		  Ellipticity         & $0.21^{+0.02}_{-0.02}$      & $0.29^{+0.04}_{-0.04}$     & $0.33^{+0.02}_{-0.02}$    \\
		  $R_p$      [arcmin] & $9.4^{+0.6}_{-0.9}$         & $13.9^{+2.1}_{-2.7}$       & $11.8^{+1.2}_{-0.6}$      \\
		  $R_p$      [pc]     & $287.1^{+27.5}_{-18.3}$     & $424.6^{+64.1}_{-45.8}$    & $360.4^{+36.7}_{-18.3}$   \\
		  reduced-$\chi^2$    & $4.51$                      & $1.87$                     & $2.75$                    \\
		{\bf Exponential profile} & & &\\    
		  Ellipticity         & $0.20^{+0.02}_{-0.02}$      & $0.26^{+0.04}_{-0.04}$     & $0.33^{+0.02}_{-0.02}$    \\
		  $R_e$      [arcmin] & $9.1^{+0.9}_{-0.6}$         & $13.3^{+2.1}_{-1.5}$       & $11.2^{+0.9}_{-0.6}$      \\
		  $R_e$      [pc]     & $277.9^{+27.5}_{-18.3}$     & $406.2^{+63.4}_{-45.8}$    & $342.1^{+27.5}_{-18.3}$   \\
		  reduced-$\chi^2$    & $4.59$                      & $1.44$                     & $4.19$                    \\
	\hline
	\end{tabular}
	}
\end{table}

%%%%%%%%%%%%%%%%%%%%%%%%%%%%%%%%%%%%%%%%%%%%%%%%%%
%%% Sec.2 %%%%%%%%%%%%%%%%%%%%%%%%%%%%%%%%%%%%%%%%
%%%%%%%%%%%%%%%%%%%%%%%%%%%%%%%%%%%%%%%%%%%%%%%%%%
\section{Intermediate- and old-age stellar populations of Carina} \label{sec2}

The Carina dSph galaxy is centered at $(\alpha_{2000},\delta_{2000})=(06^{\rm h}41^{\rm m}36.7^{\rm
s},-50^{\circ}57^{\prime}58^{\prime\prime})$ with a position angle measured North through East of~$65^{\circ}$, and is
located at a heliocentric distance, $D_{\odot}=106\pm6$~kpc~\citep{IH1995,McC2012}. In this section, we briefly
introduce the photometric and kinematic data of two stellar components in Carina dSph~\citep[][for more
details]{Bonetal2010,Monetal2014,Fabetal2016}

In this work, we use the photometric data from three different telescopes,~i.e., the CTIO 1.5~m telescope, the CTIO 4~m
Blanco telescope, and the ESO/MPG 2.2~m telescope~\citep{Bonetal2010} and the spectroscopic data from high and medium
resolution spectrographs mounted at the Very Large Telescope~\citep{Fabetal2016}. Based on photometric $V-c_{\rm
U,B,I}$ diagram, i.e. V-band magnitude versus the new photometric color index, $c_{\rm U,B,I}=(U-B)-(B-I)$, for the
individual stars~\citep{Monetal2013}, the member stars of Carina can be separated into the intermediate-age~(red clump
and subgiant stars) and the old-age~(horizontal branch and subgiant stars) stellar populations. This is possible
because $c_{\rm U,B,I}$ is strongly sensitive to stellar metallicity, especially helium and light-element
content~\citep{Bonetal2010}\footnote{Since there are two well defined star formation events
characterizing Carina stellar populations, this age separation is plausible, even though $c_{\rm U,B,I}$ is mainly
sensitive to chemical compositions. Namely, the more metal-poor population would be intrinsically older in a
stellar system.}. Figure~\ref{fig:CMD} shows the $V$~versus~$B-I$ (left panel) and $V$~versus~$c_{\rm U,B,I}$ diagram
(right panel) for the photometric observations of Carina. It is clear from the right panel of Figure~\ref{fig:CMD} that
the Carina dSph clearly has two main sub-populations.

On the other hand, performing new observations and reducing the data for the stellar spectra, and collecting
other currently available data for Carina, \citet{Fabetal2016}~performed a detailed analysis of velocity distributions
of 1096 intermediate-age and 293 old population stars. This analysis revealed that while the
intermediate-age stellar component shows a weak rotational pattern around the minor axis, the old one has a larger
line-of-sight velocity dispersion and is more extended towards the outer region. They also looked into stellar
distributions of both populations and found that the intermediate-age stars in this galaxy are centrally
concentrated, whereas the old ones are more extended in the outer region. These photometric and kinematic features are
roughly in agreement with those of similar multiple populations in Sculptor, Sextans and
Fornax~\citep{Batetal2008,Batetal2011,AmE2012}. Thus, these features of multiple stellar populations in the luminous
dSphs could be general and universal.

In what follows, using in total 6633~(3757 intermediate and 2876 old) photometric and 1389~(1096 intermediate and 293
old) spectroscopic measurements, which is spectroscopic data volume almost twice larger than
that of previous works, we investigate structural properties of stellar distributions for each stellar population and
the dark matter distribution in the Carina dSph.

\begin{figure*}
	\includegraphics[scale=0.3]{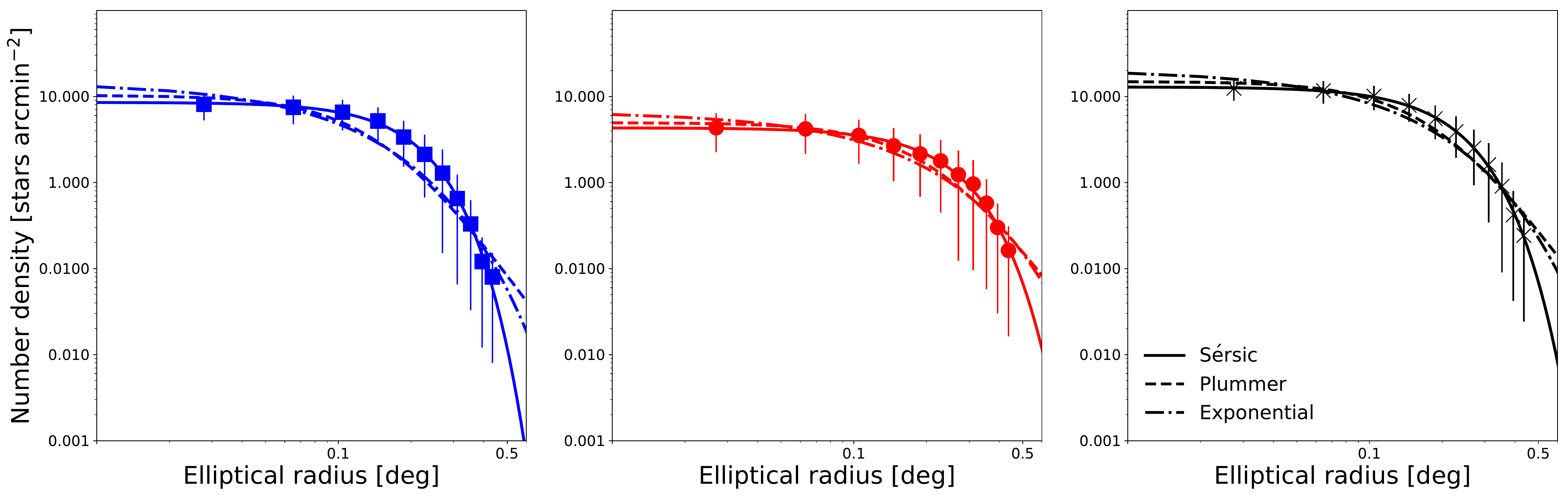}
\caption{Radial density profiles of the intermediate~(left), old~(middle), and all~(right) populations in Carina,
respectively. The points with error bars in each panel denote the observed stellar density profiles
from the photometric data. The solid, dashed and dot-dashed lines are the best-fit stellar profiles modelled by the
S\'ersic, Plummer, and the exponential profile, respectively.}
\label{RadialPro}
\end{figure*}

%%%%%%%%%%%%%%%%%%%%%%%%%%%%%%%%%%%%%%%%%%%%%%%%%%
%%% Sec.3 %%%%%%%%%%%%%%%%%%%%%%%%%%%%%%%%%%%%%%%%
\section{Two-component Mass Modeling for Carina} \label{sec3}
%%%%%%%%%%%%%%%%%%%%%%%%%%%%%%%%%%%%%%%%%%%%%%%%%%
%%% Sec.3.1 %%%%%%%%%%%%%%%%%%%%%%%%%%%%%%%%%%%%%%

\subsection{Luminous components}

In order to solve the second-order Jeans equations and set constraints on dark halo properties of the Carina dSph, we
should first obtain the structural properties of stellar populations. Here, we present briefly the procedure to estimate
their stellar profiles.

In this work, we assume that the member stars on the sky plane are distributed by the S\'ersic profile~\citep{Ser1968},
\begin{equation}
	I_{\rm Ser}(R^{\prime}) = I_{S,0}\exp\Bigl[-\Bigl(\frac{R^{\prime}}{R_s}\Bigr)^{\frac{1}{m}}\Bigr],
\label{sersic}
\end{equation}
by the Plummer profile~\citep{Plu1911}
\begin{equation}
	I_{\rm Plm}(R^{\prime}) = I_{P,0}\frac{R^2_{p}}{(R^{2}_p+R^{\prime2})^2},
\label{plummer}
\end{equation}
and by the exponential profile,
\begin{equation}
	I_{\rm Exp}(R^{\prime}) = I_{E,0}\exp\Bigl\{-\Bigl(\frac{R^{\prime}}{R_e}\Bigr)\Bigr\},
\label{exp}
\end{equation}
where $I_{(S,P,E),0}$ are the central surface densities, and ${R_{(s,p,e)}}$ denote the scale radii of each stellar
distribution. $m$ in S\'ersic profile is the S\'ersic index, which measures the curvature of the stellar profile, and
$m=1$ corresponds to the exponential profile. The projected elliptical radius, $R^{\prime}$, related to projected
sky coordinates $(\alpha,\delta)$, is given by
\begin{equation}
	R^{\prime2} = (X\cos\theta-Y\sin\theta)^2 + \Bigl[\frac{1}{1-\epsilon}(X\sin\theta+Y\cos\theta)\Bigr]^2,
\label{elliptical}
\end{equation}
\begin{equation}
	X = (\alpha-\alpha_0)\cos(\delta_0),\hspace{2mm}Y_i = \delta - \delta_0,
\end{equation}
where $(\alpha_0,\delta_0)$ is the centroid of the system, $\theta$ is the position angle of the major axis, defined
East of North on the sky, and $\epsilon$ is an ellipticity defined as $\epsilon=1-q^{\prime}$, with $q^{\prime}=b/a$
being the projected minor-to-major axial ratio of the galaxy.

In order to estimate the stellar structural parameters of Carina, we compare the radial profiles
estimated from the photometric data with those of the modelled profiles with some free parameters. First, adopting the
standard binning approach, we estimate the projected radial profiles for each population. We set the radial bins so that
a nearly equal number of stars is contained in each bin, and then derive the surface density profiles using the
number of stars contained in each bin. The points with error bars in Figure~\ref{RadialPro} show the binned profiles
estimated from the photometric data. The error bars are assumed to be the Poisson errors. To obtain the stellar
structural parameters of the assumed models by comparing with observational data, we employ a simple $\chi^2$ test,
\begin{eqnarray}
	\chi^2 = \sum^{N_{\rm bins}}_{i} \frac{[I^{\rm obs}_{i} - I^{\rm model}_{i}(O)]^2}{\varepsilon^{2}_{i}},
\label{chi2}
\end{eqnarray}
where $N_{\rm bins}$ is the number of bins, $I^{\rm obs}$ is the measured stellar surface density, $I^{\rm model}$ is
the modelled density at the same distance from the center of the system, and $\varepsilon$ is the error on $I^{\rm
obs}$. $O$ denotes the structural parameters of the assumed models, that is, $I_{(S,P,E),0}$, ${R_{(s,p,e)}}$,
$\epsilon$, and the S\'ersic index $m$. We employ the reduced $\chi^2$ statistics which takes the value of $\chi^2$
divided by the number of degrees of freedom, usually given by $N_{\rm bins}-N_{\rm parameters}$, where $N_{\rm
parameters}$ is the number of parameters.

The fitting results for the structural parameters are tabulated in Table~\ref{table1}, and Figure~\ref{RadialPro} shows
the radial profiles using the best-fit values of the parameters. Although we estimate also the best-fit value of the
parameter $I_{(S,P,E),0}$, it is not an important parameter because it should vanish when we calculate Jeans equations,
as discussed below. Thus, this parameter is not tabulated in Table~\ref{table1}. According to these results, we confirm
that the intermediate-age subcomponent is more centrally concentrated than the old one, which is in agreement
with~\citet[][see also~\citealt{Fabetal2016}]{Monetal2003}, and find that the S\'ersic profile is
the best-fit model in each population among the three stellar profile models we considered. Thus, we adopt the S\'ersic
profile as the fiducial stellar distribution of each population and incorporate it into Jeans equations.

The three-dimensional stellar density $\nu(r_{\ast})$ is obtained from the surface density $I_{\rm Ser}(R^{\prime})$ by
deprojection through the Abel transform. In the case of $m=1$, we can obtain it in the analytical form,
$\nu(r_{\ast},m=1)=S_0K_0(r_{\ast}/R_s)/(\pi R_s)$, where $K_0(x)$ is the modified Bessel function of the second kind.
On the other hand, in the case of $0.5\leq m\leq10$, there is an excellent approximation for $\nu(r_{\ast})$ derived
by~\citet{Limetal1999}:
\begin{equation}
	\nu(r_{\ast}) = \nu_0 \Bigl(\frac{r_{\ast}}{R_s}\Bigr)^{-p}\exp\Bigl[-\Bigl(\frac{r_{\ast}}{R_s}\Bigr)^{1/m}\Bigr],
\label{3D}
\end{equation}
where
\begin{eqnarray}
\nu_0 &=& \frac{I_0\Gamma(2m)}{2R_s\Gamma[(3-p)m]}, \nonumber \\
p &=& 1.0-\frac{0.6097}{m}+\frac{0.05463}{m^2}, \nonumber
\end{eqnarray}
and $\Gamma(x)=\int^{\infty}_{0}t^{x-1}e^{-t}dt$ is the gamma function.
The variable $r_{\ast}$ is expressed by the elliptical radius, $r_{\ast}^2=R^2+z^2/q^2$ in cylindrical coordinates, so
$\nu(r_{\ast})$ is constant on spheroidal shells with intrinsic axial ratio $q$ which is related to the projected axial
ratio $q^{\prime}$ and an inclination angle for the stellar system $i$ so that $q^{\prime2}=\cos^2i+q^2\sin^2i$, where
$i=90^{\circ}$ when a galaxy is edge-on and $i=0^{\circ}$ for face-on. The intrinsic axial ratio can be derived from
$q=\sqrt{q^{\prime2}-\cos^2i}/\sin i$, and thus the inclination angle is constrained by $(q^{\prime2}-\cos^2i)$ being
positive.

%%%%%%%%%%%%%%%%%%%%%%%%%%%%%%%%%%%%%%%%%%%%%%%%%%%%%%%%%%%%
%%% Sec.3.2 %%%%%%%%%%%%%%%%%%%%%%%%%%%%%%%%%%%%%%%%%%%%%%%%

\subsection{Dark components}

For the dark matter halo, we adopt a generalized Hernquist profile given by~\citet{Her1990} and~\citet{Zhao1996}, but
here we consider the non-spherical dark matter halos of dSphs
\begin{eqnarray}
&& \rho(R,z) = \rho_0 \Bigl(\frac{r}{b_{\rm halo}} \Bigr)^{-\gamma}\Bigl[1+\Bigl(\frac{r}{b_{\rm halo}} \Bigr)^{\alpha}\Bigr]^{-\frac{\beta-\gamma}{\alpha}},
 \label{DMH} \\
&& r^2=R^2+z^2/Q^2,
\label{DMH2}
\end{eqnarray}
where the six parameters are the following: $\rho_0$ and $b_{\rm halo}$ are the scale density and radius, $\alpha$ is
the sharpness parameter of the transition from the inner slope $-\gamma$ to the outer slope $-\beta$, and $Q$ is a
constant axial ratio of the dark matter halo. For $(\alpha,\beta,\gamma)=(1,3,1)$, we recover the NFW
profile~~\citep[][]{NFW1996,NFW1997} motivated by cosmological pure dark matter simulations, while
$(\alpha,\beta,\gamma)=(1.5,3,0)$ corresponds to the Burkert cored profile \citep{Bur1995}. Therefore, this dark matter
halo model enables us to explore a wide range of physically plausible dark matter profiles.

An advantage of this assumed profile is that the form of equations (\ref{DMH}) and (\ref{DMH2}) allows us to
calculate the gravitational force in a straightforward way~\citep{vanetal1994,BT2008}. By using a new variable of
integration~$\tau\equiv a^{2}_{0}(1-Q^2)[\sinh^2u_m-((1-Q^2)^{-0.5}-1)](a_0={\rm const})$ in the spheroidal
coordinate~$(u_m,v_m)$, the gravitation force can be obtained by one-dimensional integration
\begin{equation}
{\bf g} = -\nabla\Phi = -\pi GQa_0\int^{\infty}_{0} d\tau\frac{\rho(\tilde{r}^2)\nabla \tilde{r}^2}{(\tau+a^2_0)\sqrt{\tau+Q^2a^2_0}},
\end{equation}
where $G$ is the gravitational constant, and $\tilde{r}^2$ is defined by
\begin{equation}
\frac{\tilde{r}^2}{a^2_0} = \frac{R^2}{\tau+a^2_0} + \frac{z^2}{\tau+Q^2a^2_0}.
\end{equation}

%%%%%%%%%%%%%%%%%%%%%%%%%%%%%%%%%%%%%%%%%%%%%%%%%%%%%%%%%%%%
%%% Sec.3.3 %%%%%%%%%%%%%%%%%%%%%%%%%%%%%%%%%%%%%%%%%%%%%%%%

\subsection{Axisymmetric Jeans analyses}

Dwarf spheroidal galaxies are generally regarded as collisionless systems. The spatial and
velocity distributions of stars in such a dynamical system are described by its phase-space distribution function~(DF).
As the system is in dynamical equilibrium and collisionless under the smooth gravitational potential, the DF obeys the
steady-state collisionless Boltzmann equation \citep{BT2008}. Since the families of solutions satisfied by this equation
are innumerable, additional assumptions and simplifications are required. One of classical and useful ways to
alleviate this issue is to take the velocity moments of the DF. The equations derived from this approach are called
Jeans equations. In what follows, we describe these equations for the cases of axisymmetric mass distributions. This is
motivated by the fact that the luminous part of dSphs is not really spherically symmetric, nor are the shapes of dark
matter halos predicted by high-resolution $\Lambda$-dominated CDM simulations.

In the axisymmetric case, a specific but well-studied assumption is to suppose that the distribution function is of
the form $f(E,L_z)$, where $E$ and $L_z$ denote the binding energy and the angular momentum component toward the
symmetry axis, respectively. In this case, the mixed velocity moments vanish and the velocity dispersion
$(\overline{v^2_R}, \overline{v^2_{\phi}}, \overline{v^2_z})$ in cylindrical coordinates $(R,\phi,z)$ obeys
$\overline{v^2_R}=\overline{v^2_z}$ \citep[e.g.,][]{BT2008,HC2012}, that is, a velocity anisotropy parameter $\beta_z$
defined as $\beta_z=1-\overline{v^2_z}/\overline{v^2_R}$ is zero. However, $\beta_z$ could be in
general non-zero and is degenerated strongly with the dark-halo shape~\citep[][H16]{Cap2008,HC2015b}, and thus non-zero
$\beta_z$ is essential to obtain more convincing results from axisymmetric mass models. In this work, we follow the
approach of \citep{Cap2008} who assumed that the velocity ellipsoid is aligned with the $(R,\phi,z)$
coordinate and the anisotropy is constant. We note that these assumptions are supported by dark matter
simulations performed by \cite{Veretal2014}. Using the velocity anisotropy, the second-order axisymmetric Jeans
equations can be written as
\begin{eqnarray}
\overline{v^2_z} &=&  \frac{1}{\nu(R,z)}\int^{\infty}_z \nu\frac{\partial \Phi}{\partial z}dz,
\label{AGEb03}\\
\overline{v^2_{\phi}} &=& \frac{1}{1-\beta_z} \Biggl[ \overline{v^2_z} + \frac{R}{\nu}\frac{\partial(\nu\overline{v^2_z})}{\partial R} \Biggr] +
R \frac{\partial \Phi}{\partial R},
\label{AGEb04}
\end{eqnarray}
where $\nu$ is the three-dimensional stellar density and $\Phi$ is the gravitational potential dominated by dark matter.
We also assume for simplicity that the density of the tracer stars has the same orientation and symmetry as that of the
dark halo.

In order to compare second velocity moments from Jeans equations with those from the observations, we need to derive
projected second velocity moments by integrating $\overline{v^2_z}$ and $\overline{v^2_{\phi}}$ along the
line of sight. To do this, we follow the method given in~\citet[][see also~\citealt{HC2012}]{TT2006}, which took into
account the inclination of the galaxy with respect to the observer.

%%% Table 2 %%%
\begin{table*}
	\centering
	\caption{Parameter constraints with Joint and Single likelihood functions. Errors correspond to the $1\sigma$ range from our analysis.}
	\label{table2}
	\scalebox{0.7}[0.85]{
	\begin{tabular}{lcccccccccc} % four columns, alignment for each
		\hline\hline
Parameters   & $Q$ & $\log_{10}(b_{\rm halo})$  & $\log_{10}(\rho_0)$      &  $\alpha$ &   $\beta$  & $\gamma$  &  $-\log_{10}[1-\beta_{z,{\rm ALL}}]$ &  $-\log_{10}[1-\beta_{z,{\rm Int}}]$ & $-\log_{10}[1-\beta_{z,{\rm Old}}]$   &   $i$  \\
             &     & [pc]                       &  [$M_{\odot}$ pc$^{-3}$] &           &            &           &                                      &                                      &                                       &  [deg] \\            
		\hline
{\bf Joint}        & $1.23^{+0.51}_{-0.52}$ & $4.11^{+0.55}_{-0.58}$ & $-1.70^{+0.34}_{-0.52}$ & $1.85^{+0.74}_{-0.77}$ & $6.27^{+2.42}_{-2.35}$ & $0.28^{+0.23}_{-0.17}$ &---& $0.43^{+0.18}_{-0.22}$ & $0.40^{+0.19}_{-0.22}$ & $71.35^{+12.28}_{-12.66}$\\
{\bf Single}       & $1.34^{+0.42}_{-0.55}$ & $4.21^{+0.54}_{-0.61}$ & $-1.69^{+0.26}_{-0.42}$ & $2.24^{+0.52}_{-0.67}$ & $6.68^{+2.20}_{-2.40}$ & $0.21^{+0.19}_{-0.14}$ & $0.46^{+0.18}_{-0.23}$ &---&---& $72.89^{+10.89}_{-13.38}$\\
	\hline
	\end{tabular}
	}
\end{table*}

\begin{figure*}
	\includegraphics[scale=0.06]{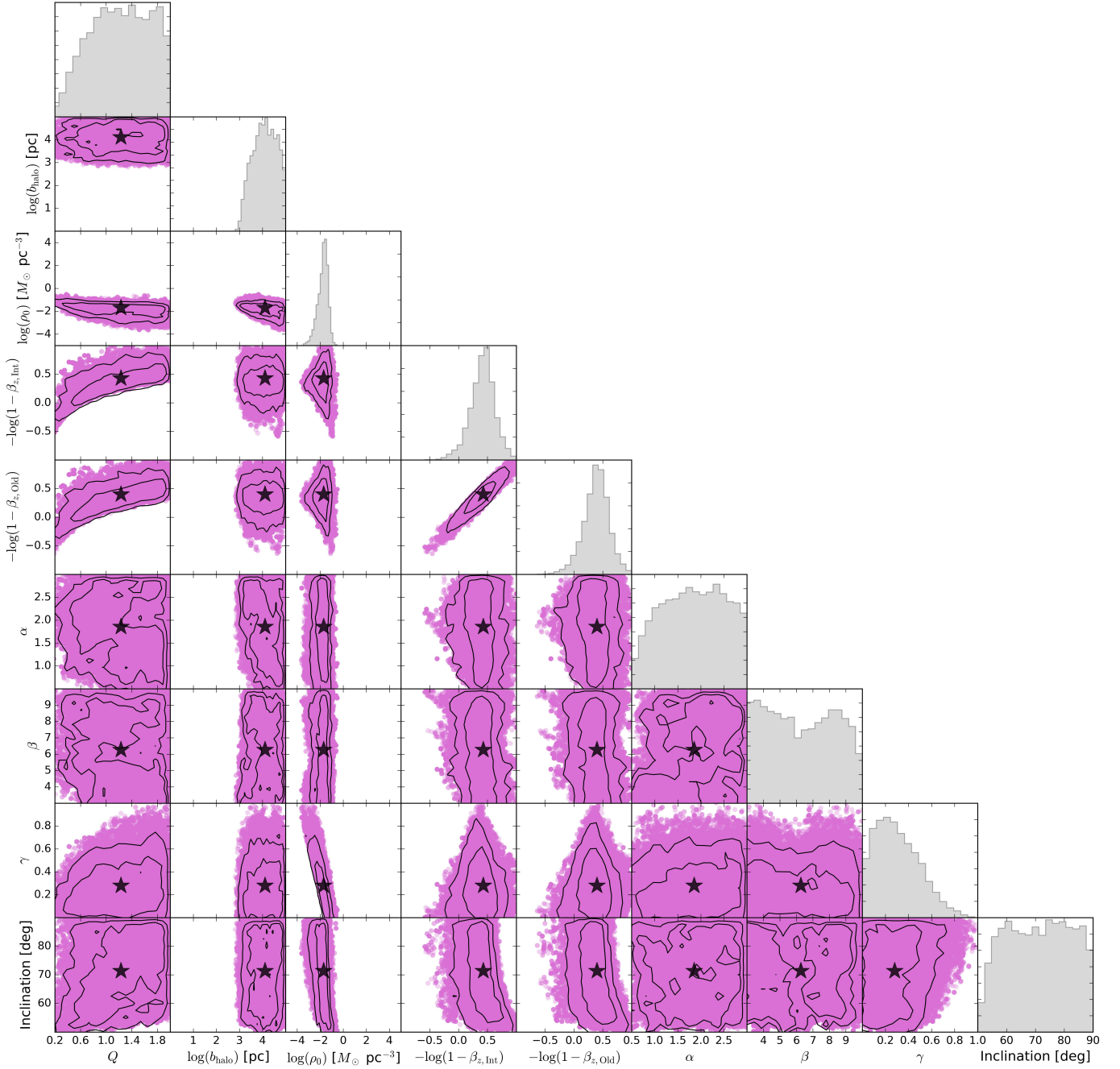}
\caption{Posterior distributions for the fitting parameters calculated from the joint likelihood analysis.
The star in each panel denotes the median value of each parameter. The black lines represent the
$1\sigma$ and $2\sigma$ regions. For some parameters, especially $\alpha$ and $\beta$, the distributions are not well
constrained.}
\label{PDF}
\end{figure*}

%%%%%%%%%%%%%%%%%%%%%%%%%%%%%%%%%%%%%%%%%%%%%%%%%%%%%%%%%%%%
%%% Sec.4 %%%%%%%%%%%%%%%%%%%%%%%%%%%%%%%%%%%%%%%%%%%%%%%%%%
%%%%%%%%%%%%%%%%%%%%%%%%%%%%%%%%%%%%%%%%%%%%%%%%%%%%%%%%%%%%

\section{Fitting Procedure}
%%% Sec.4.1 %%%%%%%%%%%%%%%%%%%%%%%%%%%%%%%%%%%%%%%%%%%%%%%%%%

\subsection{Likelihood function}

We consider different stellar populations within the same gravitational potential originating from a dark halo, but
each population shows a different spatial and velocity distribution. Therefore, we set the same likelihood function
for each population and combine them to obtain tighter constraints on the dark halo parameters.

For each stellar population $k=\{{\rm Int, Old}\}$~(where Int and Old mean intermediate- and old-age populations,
respectively), we assume that the line-of-sight velocity distribution is Gaussian and centered on the systemic velocity
of the galaxy $\langle u \rangle$. Given that the total number of member stars for each population is $N^{k}$, and the
$i$th star has the measured line-of-sight velocity $u_i\pm\delta_{u,i}$ at the sky plane coordinates~$(x_i,y_i)$, the
likelihood function for each population $k$ is constructed as
\begin{equation}
{\cal L}^{k} = \prod^{N^{k}}_{i=1}\frac{1}{(2\pi)^{1/2}[(\delta^k_{u,i})^2 + (\sigma^k_i)^2]^{1/2}}\exp\Bigl[-\frac{1}{2}\frac{(u^k_i-\langle u^k \rangle)^2}{(\delta^k_{u,i})^2 + (\sigma^k_i)^2} \Bigr],
\end{equation}
where $\sigma^k_i$ is the theoretical line-of-sight velocity dispersions at~$(x_i,y_i)$ specified by model
parameters~(as described below) and derived from the Jeans equations. Then, combining the intermediate and old
populations, we obtain the ``joint'' likelihood function
\begin{equation}
{\cal L}_{\rm Joint} = \prod_{k}{\cal L}^{k} = {\cal L}^{\rm Int} \times {\cal L}^{\rm Old}.
\label{eq:JLF}
\end{equation}

We also perform the fitting analysis for the whole stellar sample using ``single'' likelihood function, which means
that we do not separate between the intermediate- and old-age stellar populations, to compare with the results from
the joint likelihood function.

%%%%%%%%%%%%%%%%%%%%%%%%%%%%%%%%%%%%%%%%%%%%%%%%%%%%%%%%%%%%
%%% Sec.4.2 %%%%%%%%%%%%%%%%%%%%%%%%%%%%%%%%%%%%%%%%%%%%%%%%

\subsection{Model parameters}

For the case of axisymmetric models, the model parameters are the six parameters of the dark halo and three parameters
of the stellar properties in the axisymmetric case, for which we adopt the uniform/log-uniform priors. Here we
describe these parameters.

Since we assume that the different stellar populations move in the gravitational
potential of a dark halo with the generalized Hernquist density profile, there are six free parameters of the dark
halo profile:~(1) the axial ratio of dark halo~$Q$, (2) the scale radius~$b_{\rm halo}$, (3) the scale
density~$\rho_0$, (4) the sharpness of the transition from the inner to the outer density slope~$\alpha$, (5) the outer
density slope~$\beta$, and (6) the inner density slope~$\gamma$. For the stellar component, we consider different
velocity anisotropy parameters, whilst we assume that the inclination of each stellar population is identical. Thus,
there are three free parameters:~(7) the velocity anisotropy for th intermediate-age population~$\beta_{z,{\rm Int}}$,
(8) that for old-age one~$\beta_{z,{\rm Old}}$, and (9) the inclination angle~$i$. For this total of nine parameters,
the prior ranges of each parameter are
\begin{description}
\item[(1)] $0.1 \leq Q \leq 2.0$;\\
\item[(2)] $0 \leq \log_{10}[b_{\rm halo}/{\rm pc}] \leq +5$;\\
\item[(3)] $-5 \leq \log_{10}[\rho_0/(M_{\odot}{\rm pc}^{-3})] \leq +5$;\\
\item[(4)] $0.5 \leq \alpha \leq 3$;\\
\item[(5)] $3 \leq \beta \leq 10$;\\
\item[(6)] $0 \leq \gamma \leq 1.2$;\\
\item[(7)] $-1 \leq -\log_{10}[1-\beta_{z,{\rm Int}}] < +1$;\\
\item[(8)] $-1 \leq -\log_{10}[1-\beta_{z,{\rm Old}}] < +1$;\\
\item[(9)] $\cos^{-1}(q^{\prime}) \leq i \leq 90^{\circ}$.\\
\end{description}
As we described in Section~3.1, the lower limit of the inclination angle~$i$ is confined by $q^{\prime 2} -\cos^2i>0$.
Besides the above free parameters, the systemic velocity $\langle u^k \rangle$ of the system is also included as a free
parameter with a flat prior. When we estimate these parameters using a single likelihood function, the number of free
parameters is eight because we can then consider only one stellar velocity anisotropy parameter.

In order to explore the large parameter space efficiently, we adopt Markov Chain Monte Carlo (MCMC) techniques,
based on Bayesian parameter inference, with the standard Metropolis-Hasting algorithm \citep{Metetal1953,Has1970}. We
take several post-processing steps (burn-in step, the sampling step and length of the chain) to generate independent
samples that can avoid the influence of the initial conditions, and then we obtain the posterior probability
distribution function (PDF) of the set of free parameters. By calculating the percentiles of these PDFs, we are able to
compute credible intervals for each parameter in a straightforward way.

%%% Sec.5 %%%%%%%%%%%%%%%%%%%%%%%%%%%%%%%%%%%%%%%%%%%%%%%%

\section{Results}

In this section, we present the results of the MCMC fitting analysis for two-stellar components in Carina with nine
free parameters, and compute the median and credible interval values from the resulting PDF.

Figure~\ref{PDF} displays the posterior PDFs derived from the MCMC procedure with a joint likelihood
function~(\ref{eq:JLF}). Relatively, the halo parameters $b_{\rm halo}$ and $\rho_{0}$ are better constrained than
the other parameters. The PDFs for the inner slope of dark matter density, $\gamma$, are more widely spread than for the
above two parameters, but still they indicate that the data show preference for shallower cuspy
or cored dark matter density profile in Carina. On the other hand, $(Q,\alpha,\beta,i)$ are so widely distributed in
these parameter spaces that it is difficult to confine the parameter distributions of these. In particular, the
axis ratio of the dark halo has a strong degeneracy with the stellar velocity anisotropy, $\beta_z$. As
discussed in several papers~\citep[][and HC15]{Cap2008,BHB2013}, the change of
these has a similar effect on the profiles of the line-of-sight velocity dispersions. In
addition, there also exists a degeneracy between $\beta_z$ for the intermediate and the old stellar
population, even though both parameters have the peak of PDF at $-\log_{10}[1-\beta_z]\sim0.4$ which means that
the stellar velocity distributions for both populations in Carina are similar and are likely to possess
velocity dispersion more strongly biased in the $R$ direction than in the $z$ direction.

\begin{figure}
	\includegraphics[scale=0.38]{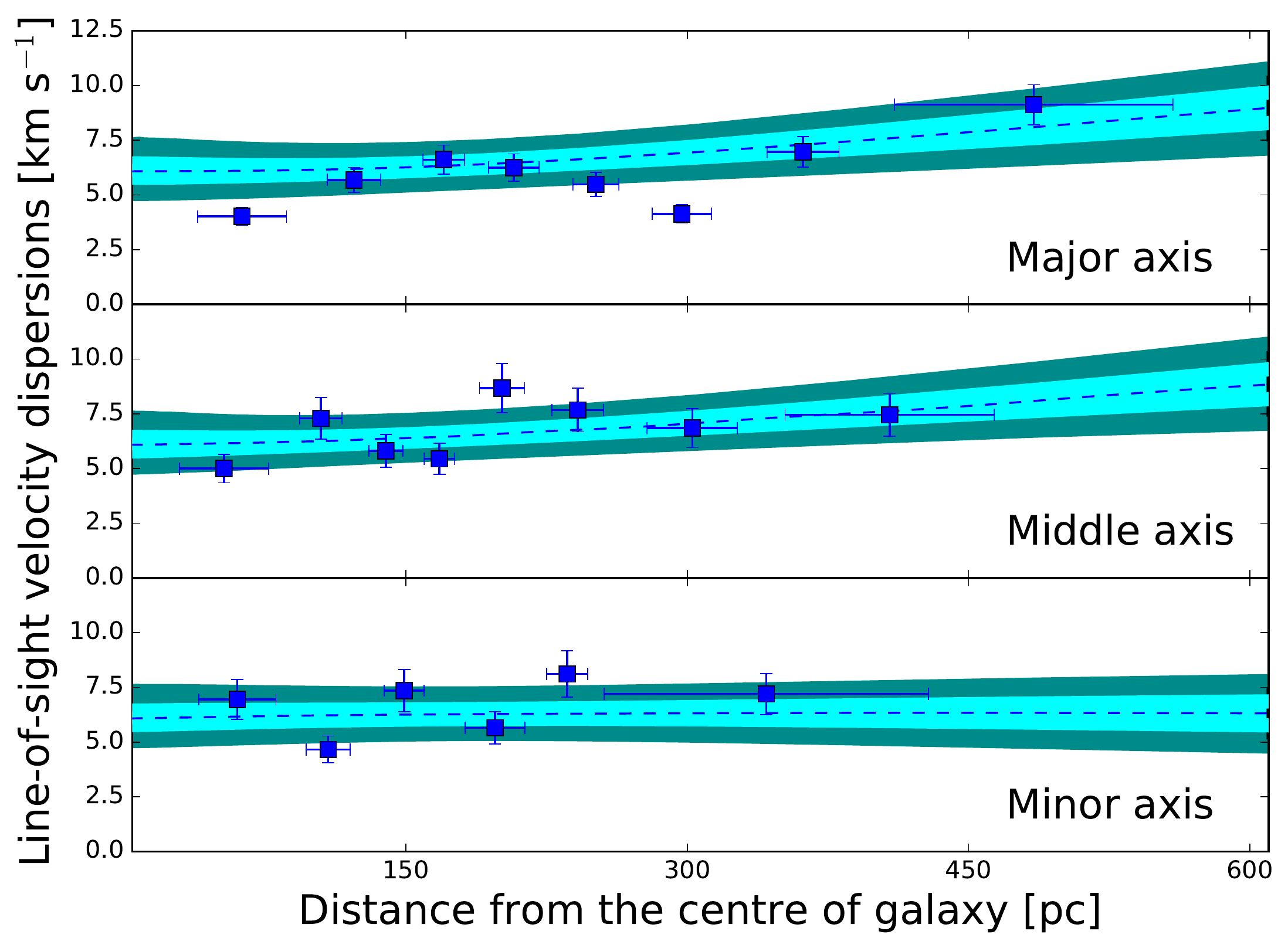}
	\includegraphics[scale=0.38]{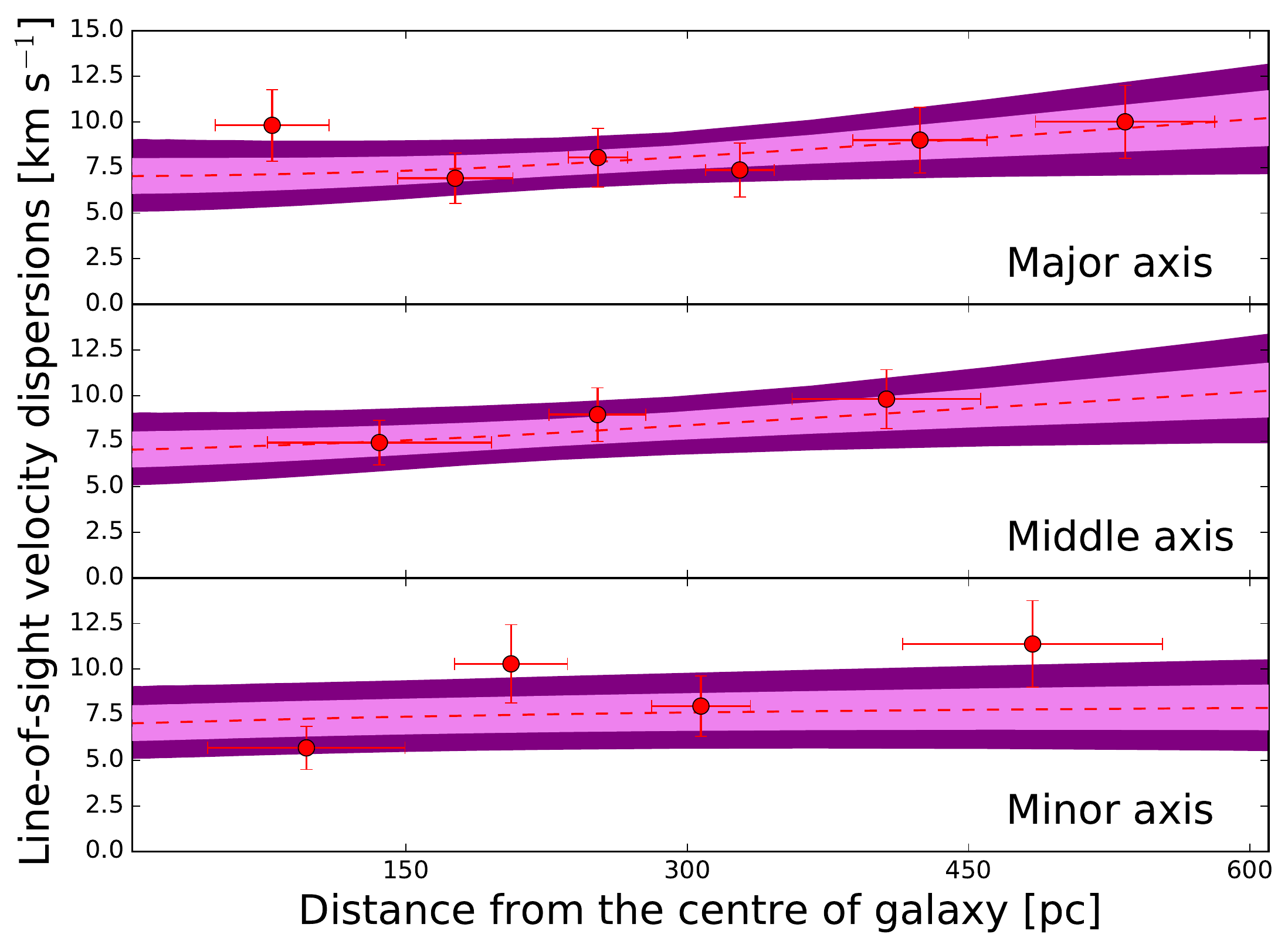}
\caption{Profiles of the line-of-sight velocity dispersions} along the major (first row), middle
(second row) and minor (third row) axis for the intermediate- (top panel) and old-age (bottom panel) population. The
colour points with error bars in each panel denote observed velocity dispersions. The dashed lines are median values of
the models and the light and dark shaded regions encompass the 68 per cent and 95 per cent confidence levels from the
results of the unbinned MCMC analysis. The methods for generating the binned profiles are described in the main text.
\label{LOS}
\end{figure}

To confirm whether our unbinned analysis can reproduce the observed kinematics of Carina, we calculate profiles of
the line-of-sight velocity dispersions for both populations using fitting parameters in our models
and the observational data. 
To estimate these profiles from the models, we obtain the output parameters of their MCMC chains, 
and then we compute the line-of-sight velocity dispersion profiles through the axisymmetric Jeans equations for all 
cases of the output parameter sets. Finally, we calculate median, 15.87th and 84.13th percentiles~(i.e. 68~per~cent 
confidence level), and 2.28th and 97.72nd percentiles (i.e. 95~per~cent confidence level) of them from all parameter sets.
On the other hand, in order to obtain these binned profiles from the observed
line-of-sight velocity data, we adopt the common way of using binning profiles. Firstly, assuming an axisymmetric
stellar distribution adopted in this work, the projected positions $(x,y)$ of the line-of-sight velocity data are
folded into the first quadrant. Secondly, we transform the projected stellar distribution in the first quadrant to
two-dimensional polar coordinates~$(r,\theta)$, where $\theta=0^{\circ}$ is set along the major axis of the stellar
distribution, and then divide this into three areas: $\theta=0^{\circ}-30^{\circ}$, $30^{\circ}-60^{\circ}$, and
$60^{\circ}-90^{\circ}$, respectively. For convenience, the first azimuthal region~($\theta=0^{\circ}$ and
$30^{\circ}$) is referred to as a major axis, the second one as a middle axis, and third one as a minor axis. 
Finally, for each area, we radially divide stars into bins so that a nearly equal number of stars is contained in each bin. 
Figure~\ref{LOS} shows this quantity along the projected major, minor and middle axis for
the intermediate and old stellar populations in Carina. 
The dashed line and the shaded region in each panel denote the median value and confidence levels (light: 68 per~cent,
dark: 95 per~cent) of the resultant unbinned MCMC analysis of the model profiles, whereas the points with error bars
are binned second velocity moments calculated from the observed data with the above method. As shown in this figure,
the profiles calculated from the results of our MCMC analysis are in good agreement with binned profiles along each
axis, even though we do not perform any binned fitting analysis. Furthermore, it is found that all 
second moment curves are almost flat, irrespective of the stellar population and the axis direction. These flat profiles 
are similar to those found in previous works even assuming spherically symmetric stellar distributions~\citep[e.g.,][]{Waletal2009c}.

Table~\ref{table2} tabulates the results of MCMC fitting for the joint~(first row) and the single~(second row)
likelihood analysis\footnote{In addition, we perform the MCMC fitting for spherical mass models to compare with 
our axisymmetric mass models. For the spherical mass models here, we assume two cases of models:
one with the axial ratios of both dark~$(Q)$ and luminous~$(q^{\prime})$ components equal to unity, and another such that the dark halo 
is assumed to be spherical $(Q=1)$, but the stellar distributions are not spherical $(q^{\prime}\neq1)$.
Then, using the results of the MCMC fitting, we estimate a Bayes factor which is the ratio of the mean posterior 
distribution of the axisymmetric to spherical symmetric mass models.
The resulting Bayes factor in the case of $(Q=1,q^{\prime}=1)$ is 450.9, and that in the case of 
$(Q=1,q^{\prime}\neq1)$ is 37.8, respectively.
It is clearly found that  our axisymmetric mass models yield a relatively better fit than any spherical ones.}. 
We show the median and $1\sigma$~(68~per~cent) confidence intervals of the free parameters, which
correspond to the 50th~(median), 16th~(lower error) and 84th~(upper error) percentiles of the posterior PDFs,
respectively. $\beta_{\rm ALL}$ denotes the velocity anisotropy for all stars, and thus this fitting result in shown
only in the second row. Comparing the results of the joint and the single likelihood analysis, we find that there is no
significant difference in both the median and the confidence intervals for all parameters. This is
because in the joint analysis the number of stars is probably not large enough to obtain stronger
limits on the dark halo parameters than in the case of the single analysis. In order to set constraints on the dark halo
more robustly, the number of observed stars in the kinematic sample is absolutely essential. Nevertheless, the joint
analysis enables us to investigate velocity anisotropy profiles separately for each of the components~(see
Section~6.3).

%%% Sec.6 %%%%%%%%%%%%%%%%%%%%%%%%%%%%%%%%%%%%%%%%%%%%%%%%
\section{Discussion}
%%%%%%%%%%%%%%%%%%%%%%%%%%%%%%%%%%%%%%%%%%%%%%%%%%%%%%%%%%%%
%%% Sec.6.1 %%%%%%%%%%%%%%%%%%%%%%%%%%%%%%%%%%%%%%%%%%%%%%%%

\subsection{Astrophysics factors for Carina}

The Galactic dSph galaxies are ideal targets for constraining particle candidates for dark matter through
indirect searches utilizing $\gamma$-rays or X-rays originating from dark matter annihilations and decays. This is
because these galaxies possess large dark matter content, are located at relative proximity, and have low astrophysical
foregrounds of $\gamma$-rays and X-rays. The $\gamma$-ray and X-ray flux can be derived by the annihilation cross
section or decay rate which estimates how dark matter particles transform into standard model particles (the so-called
particle physics factor) and line-of-sight integrals over the dark matter distribution within the system~(the so-called
astrophysics factor). In particular, the astrophysics factor depends largely on the annihilation and decay fluxes.
Therefore, in order to obtain robust limits on particle candidates for dark matter, accurate understanding of the dark
matter distribution in dSphs is of crucial importance. So far, some previous works have already evaluated the
astrophysics factors for Carina dSph assuming spherical
distribution~\citep[e.g.,][]{Acketal2014,Geretal2015,Bonetal2015} and axisymmetric
distribution~\citep[H16,][]{Kloetal2017} of dark matter density. However, here we calculate for the first time the
astrophysics factors for Carina using the fitting results from the joint likelihood analysis of multiple stellar
components and assuming more general dark halo models which include non-sphericity and the generalized
Hernquist density profile. In this section, we estimate the astrophysics factors from our results and compare them with
those from previous studies. To do this adequately, we use only the factors integrated within a fixed integration
angle $0.5^{\circ}$.

%%% Table 3 %%%
\begin{table}
	\centering
	\caption{Comparison of $J$ and $D$ values integrated within $0.5^{\circ}$ calculated from this work and taken from the previous works. The values of error correspond $1\sigma$ uncertainties. The units of $J_{0.5}$ and $D_{0.5}$ are [GeV$^2$~cm$^{-5}$] and [GeV~cm$^{-2}$], respectively.}
	\label{table3}
		\scalebox{0.65}[0.8]{
	\begin{tabular}{cccc} % four columns, alignment for each
		\hline\hline
		                     & This work     & \citet[][H16]{Hayetal2016} & \citet{Geretal2015} \\
         DM Models           & Axisymmetric &  Axisymmetric       &      Spherical       \\
                             \hline
        $\log_{10}[J_{0.5}]$ & $19.08^{+0.54}_{-0.63}$ & $17.97^{+0.46}_{-0.28}$ & $17.87^{+0.10}_{-0.09}$ \\
        $\log_{10}[D_{0.5}]$ & $19.07^{+0.29}_{-0.63}$ & $18.19^{+0.26}_{-0.25}$ & $17.90^{+0.17}_{-0.16}$ \\
	\hline
	\end{tabular}
	}
\end{table}

The astrophysics factors for dark matter annihilation and decay are given by
\begin{eqnarray}
J &=& \int_{\Delta\Omega}\int_{\rm los}d\ell d\Omega \hspace{1mm}\rho^2(\ell,\Omega) \hspace{3.5mm} {\rm [annihilation]}, \label{eq:J}\\
D &=& \int_{\Delta\Omega}\int_{\rm los}d\ell d\Omega \hspace{1mm}\rho(\ell,\Omega) \hspace{4.9mm} {\rm [decay]}, \label{eq:D}
\end{eqnarray}
which are called $J$~factor and $D$~factor~\citep[e.g.,][]{Gunetal1978,Beretal1998}. These factors correspond to the
line-of-sight integral of the square of dark matter density for annihilation and the dark matter density for
decay, respectively, within the solid angle $\Delta\Omega$. 
To estimate these factors for axisymmetric dark matter distributions, the output parameters~$(Q,b_{\rm halo},\rho_0,\alpha,\beta,\gamma,i)$ from
their MCMC chains are used to calculate the dark matter annihilation $J$-factor and decaying dark matter $D$-factor integrated within 
$0.5^{\circ}$ solid angle using equations~(\ref{eq:J}) and (\ref{eq:D}).
We also calculate the median and $1\sigma$ uncertainties of $J$- and $D$-factors using the marginalized PDFs of these factors.
The methods are described in more detail in~Section~2 of H16.

\begin{figure}
	\includegraphics[scale=0.45]{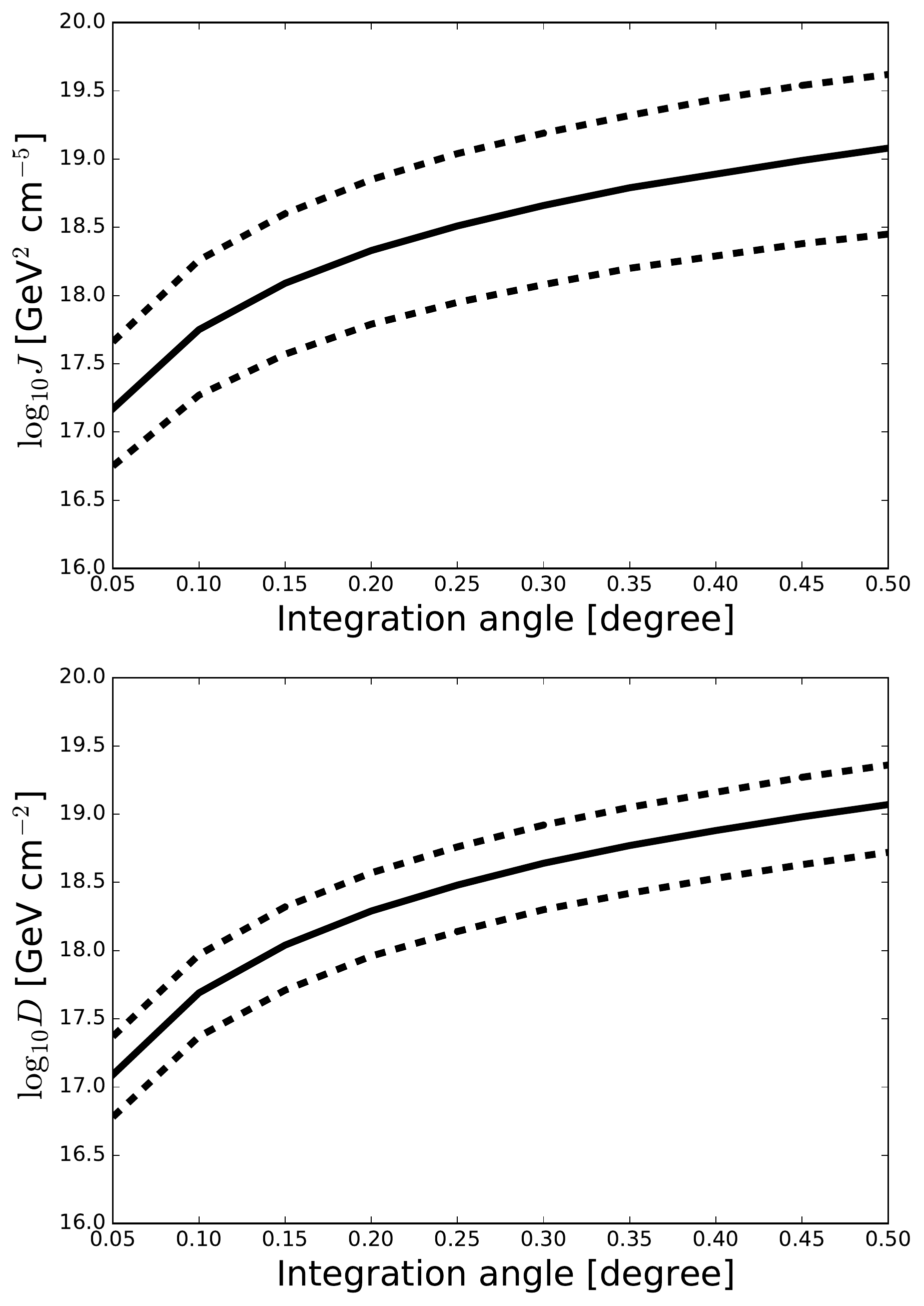}
    \caption{Expected emission profiles for annihilation~(top panel) and decay~(bottom panel). The solid line in each
panel denotes the median profile, and the dashed line corresponds to the $\pm1\sigma$ distribution.}
\label{JDprofile}
\end{figure}

Table~\ref{table3} shows a comparison of the $J$ and $D$ values integrated within $0.5^{\circ}$ solid angle of our
results with those of previous works. It is clear from this table that there are differences in the median values and
the uncertainties of the $J$- and $D$-factor between our estimates and other studies. The reasons for these differences
are not only the effects of non-sphericity but also the size of the kinematic data sample and the number of free
parameters in the assumed dark matter density profile. Since H16 have already discussed the differences in the
astrophysics factors between their axisymmetric mass models and the spherical ones from~\citet{Geretal2015}, in what
follows we focus on the differences between axisymmetric models in this work and in H16.

Naturally, the main reason why the $J$ and $D$ uncertainties from this work are larger than those from H16 would be the
difference in the number of fitting parameters of dark halo profiles. H16 assumed that the outer slope of dark matter
profiles is $\rho\propto r^{-3}$ consistently and the parameter of the transition from the inner to the outer density
slope has a constant value, so that the total number of dark halo parameters was four. On the other hand, the dark
matter profile we adopt here is a generalized Hernquist profile which takes into account the outer slope and the
curvature of the density profile as free parameters, and thus we have six parameters for the dark halo in total. As for
the difference in the median values, it can be understood by the fitting results for the scale length~$(b_{\rm halo})$
and the scale density~$(\rho_0)$ of the dark halo profile. While H16 obtained $\log_{10}(b_{\rm halo}/{\rm
[pc]})=3.5^{+0.7}_{-0.6}$ and $\log_{10}(\rho_0/[M_{\odot}{\rm pc}^{-3}])=-2.2^{+1.0}_{-0.9}$ from their MCMC fitting
procedure, our fitting result from the joint likelihood analysis gives $\log_{10}(b_{\rm halo}/{\rm
[pc]})=4.1^{+0.6}_{-0.6}$ and $\log_{10}(\rho_0/[M_{\odot}{\rm pc}^{-3}])=-1.7^{+0.3}_{-0.5}$. Therefore, the value of
the scale length (the scale density) estimated from our results is about a factor of four (three) larger than that from
H16, and thus the $J$ and $D$ factors in this study should be higher than the values in the previous axisymmetric work.
The reason for the difference in these dark halo parameters could be the different data volume of the stellar kinematic
sample. The number of available velocity data for Carina was 776 stars in previous works, whereas this
work utilizes 1389 stars to obtain limits on dark matter distribution. Thus, it is not surprising that fitting
results in this work are not consistent with those in the others.

Furthermore, we find that our evaluated $J$- and $D$-factors for Carina have higher values in comparison with those for
the other Milky Way dwarf satellites. Several known Galactic dwarf galaxies such as Draco, Ursa~Minor,
Coma~Berenices and Ursa~Major~II possess large mean $J$- and $D$-factors~\citep[e.g.,][H16]{Acketal2015,Geretal2015}.
For instance, Draco, which might have the highest $J$ and $D$ value among classical dwarfs, has
$\log_{10}J=19.09^{+0.39}_{-0.36}$~GeV$^{2}$~cm$^{-5}$ and $\log_{10}D=18.84^{+0.23}_{-0.21}$~GeV~cm$^{-2}$~as given by
H16. We therefore suggest that the Carina dSph becomes one of the most promising detectable
target among the classical dwarf galaxies for an indirect search of dark matter annihilation and decay.

Finally, to facilitate the evaluation of a survey design using $\gamma$-ray or X-ray observation for Carina, we quantify
the spatial extent of the emission from dark matter annihilation and decay. The angular distributions
of emission can provide useful information about how large areas are required in optimal observations.
Moreover, the detection of a spatially varying annihilation or decay emission signal is directly linked to the
possibility of revealing the nature of the dark matter halo. Figure~\ref{JDprofile} displays the median value and
$1\sigma$ confidence intervals of the $J$- and $D$-factor as a function of the integration solid angle. We calculate
astrophysical factors extended to $0.5^{\circ}$ integration angle, because the outermost observed member star in
Carina is located about $0.5^{\circ}$ from the centre of the galaxy. From this figure, we find that beyond the
integration angle~$\sim0.1^{\circ}$, the $J$ and $D$ values increase moderately, thereby suggesting that to
preform the optimal $\gamma$-ray or X-ray observation of Carina, the observational area larger than $0.1^{\circ}$ would be required.

%%%%%%%%%%%%%%%%%%%%%%%%%%%%%%%%%%%%%%%%%%%%%%%%%%%%%%%%%%%%
%%% Sec.6.2 %%%%%%%%%%%%%%%%%%%%%%%%%%%%%%%%%%%%%%%%%%%%%%%%
\subsection{Dark matter density profile}

Using the fitting results for dark matter density parameters, we present the inferred dark matter
density profile of Carina. In addition, we also calculate the density profile given by \citet[][hereafter
GS15]{Geretal2015} for comparison. Both dark halo profiles were modelled with the same free parameters, except for
the difference in non-sphericity and thus we can compare between their density profiles reliably.

\begin{figure}
	\includegraphics[scale=0.45]{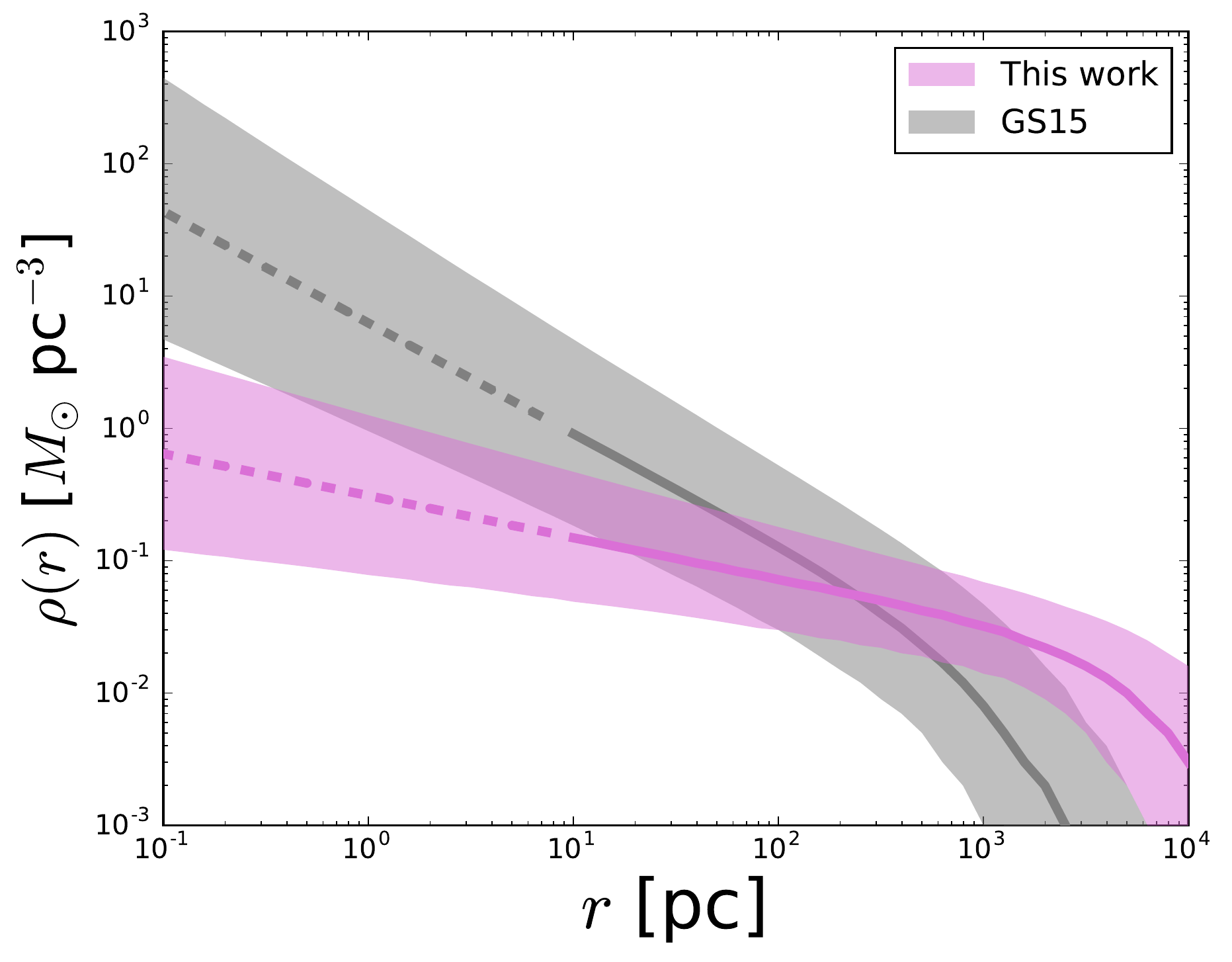}
\caption{Derived dark matter density profile from our Jeans analysis (purple) and the results of
GS15~(gray). The dashed lines denote smaller radii than the positions of the innermost observed member stars, so
the density profile within the region is less certain. On the other hand, the solid lines represent the
reliable density profiles. The shaded regions denote $\pm1\sigma$ intervals.}
\label{DMPROFILE}
\end{figure}

Figure~\ref{DMPROFILE} shows the dark matter density profile estimated from our
joint analysis~(purple) and from GS15~(gray). For our results, we compute the dark matter density profile along its
major axis. On the other hand, for GS15, we calculate the density profile using the median and $\pm1\sigma$ values of
dark halo parameters listed in table~4 of their paper. We find from this figure that although the density
distribution in the inner region is not a robust result and has large uncertainties, the inner density slope
from this work favors shallower values than that from GS15. Moreover, our estimated dark halo size, which corresponds
to the radius of the transition from the inner to the outer slope, is larger than that from the previous spherical
work. This means that in our case the dark matter would be more widely distributed, thereby implying that the
astrophysics factors can have large values even though the preferred inner dark matter density is shallower.

%%%%%%%%%%%%%%%%%%%%%%%%%%%%%%%%%%%%%%%%%%%%%%%%%%%%%%%%%%%%
%%% Sec.6.3 %%%%%%%%%%%%%%%%%%%%%%%%%%%%%%%%%%%%%%%%%%%%%%%%
\subsection{Velocity anisotropy}

\begin{figure}
	\includegraphics[scale=0.45]{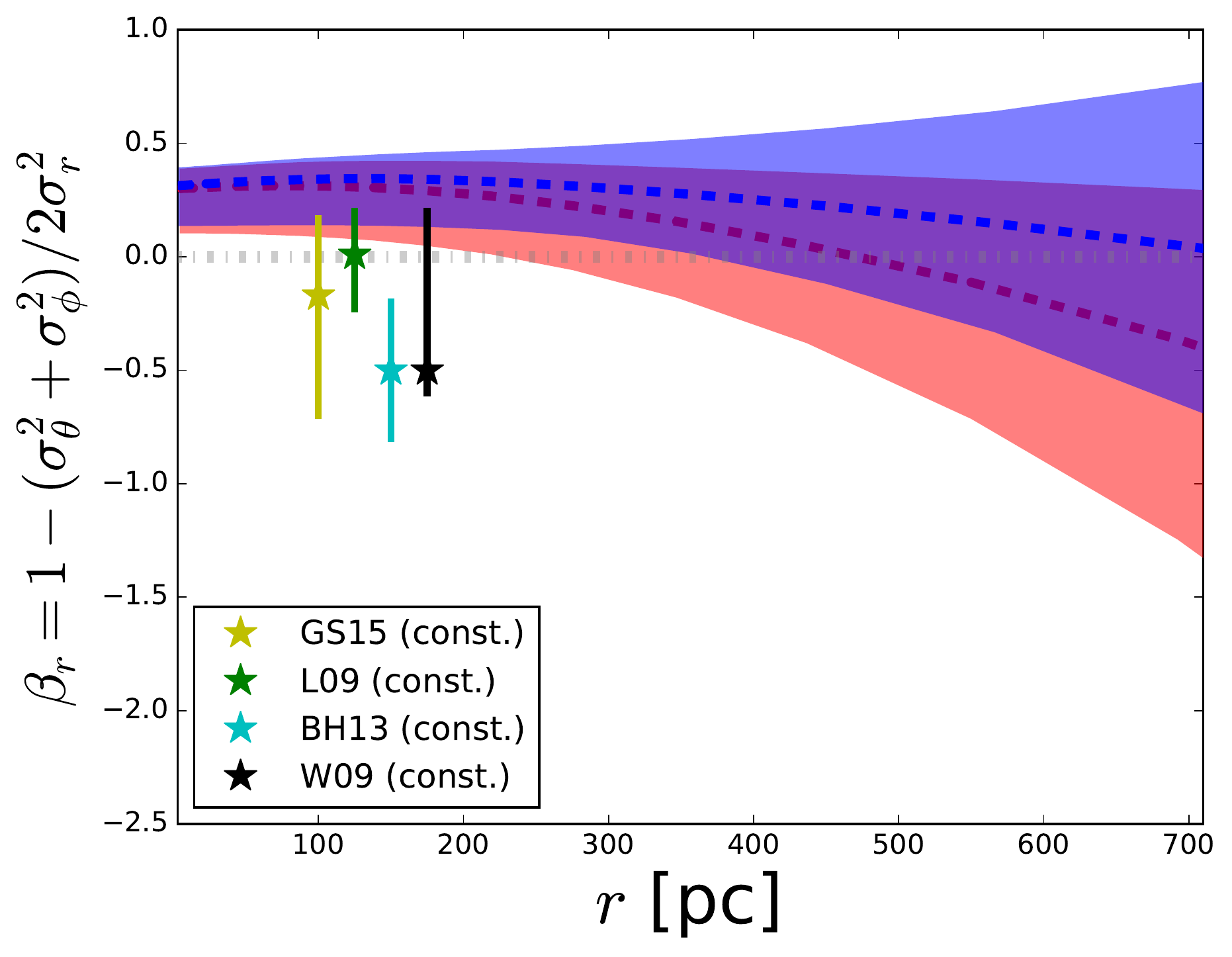}
\caption{The profiles of the velocity anisotropy $\beta_r$ for the intermediate~(blue) and old~(red) age stellar
populations. The dashed lines and shaded regions denote the median values and the $\pm1\sigma$ intervals, respectively.
The gray dash-dotted line corresponds to the isotropic value, i.e., $\beta_r=0$. The colored stars with error bars are
the constant velocity anisotropy values given by \citet[][black star]{Waletal2009c}, \citet[][green]{Lokas2009},
\citet[][cyan]{BH2013}, and \citet[][yellow]{Geretal2015}, respectively.}
\label{ANI}
\end{figure}

In general, axisymmetric models are to some extent capable of taking into account the velocity anisotropy that
depends on the spatial coordinates. This is important because revealing the stellar velocity anisotropy provides
critical information to understand the dynamical evolution history and the dark matter distribution in the system. In
order to compare with previous results, we transform the second velocity moments in cylindrical coordinates to spherical
coordinates and estimate the stellar velocity anisotropy in a spherical model which is given by
\begin{equation}
	\beta_r = 1 - \frac{\sigma^2_{\theta}+\sigma^2_{\phi}}{2\sigma^2_{r}},
\label{VESPH}
\end{equation}
where $\sigma^2_{\theta} = \overline{v^2_{\theta}}$, $\sigma^2_{\phi}=\overline{v^2_{\phi}} - \overline{v}_{\phi}^2$,
and $\sigma^2_{r} = \overline{v^2_{r}}$ in spherical coordinates~$(r,\theta,\phi)$. Figure~\ref{ANI} displays the
velocity anisotropy profiles calculated from our axisymmetric models for intermediate~(blue) and old~(red) age stellar
populations. The dashed lines denote median values of the quantity, and the shaded regions correspond to their $1\sigma$
confidence intervals. If a galaxy has an isotropic velocity ellipsoid for stars, $\beta_r$ is equal to zero. We find
from this figure that the common assumptions of anisotropy constant with radius or equal to zero would be unrealistic.
Moreover, both populations are radially anisotropic at small radii. Then, the velocity anisotropy of the intermediate
population decreases gradually to isotropic with increasing radius, while that of old one becomes moderately
tangentially anisotropic outwards, even though there are large uncertainties especially in outer parts of the system.

Figure~\ref{ANI} also shows the values of the constant velocity anisotropy estimated by previous works assuming
spherical symmetry taken from~\citet[][black star]{Waletal2009c}, \citet[][green]{Lokas2009}, \citet[][cyan]{BH2013},
and \citet[][yellow]{Geretal2015}, respectively. We note that these values of velocity anisotropy do not depend
on the radius (i.e. $x$-axis in this figure), since these are ``constant'' values of $\beta_r$.
\citet{Waletal2009c} used single-component spherical Jeans equations to model line-of-sight velocity dispersions
of the classical dSphs, and they found that Carina dSph favors a mild tangential velocity bias for any dark
halo mass profiles assumed. \citet[][]{Lokas2009} also solved spherical Jeans equations to reproduce the projected
higher-order velocity moment~(kurtosis) as well as second moment, and showed that an almost isotropic
stellar velocity distribution is preferred $(\beta_r\sim0.01)$. \citet{Geretal2015} performed dynamical analysis similar
to \citet{Waletal2009c}, with a difference in the assumed dark halo profiles and likelihood functions. They found
slightly tangentially biased velocity anisotropy. On the other hand, \citet[][]{BH2013} implemented single-component
orbit-based Schwarzschild models and computed line-of-sight velocity dispersions and kurtoses of the luminous dSphs.
From their analysis, the velocity anisotropy of Carina is clearly tangential.

Broadly, the velocity anisotropy profiles calculated from our analysis are consistent with the single component
analyses from the previous works in the outer parts. However, our axisymmetric models assume a constant $\beta_z$ for
simplicity and thereby our calculated $\beta_r$ is strongly limited. In order to investigate the detailed
profile of the velocity anisotropy, we need more realistic dynamical models for dSphs, relaxing the assumption of a
constant $\beta_z$ and more stellar kinematic data in the future.

The feature that each population has more tangential (less radial) velocity anisotropy at larger
radii, implies a vestige of tidal disturbances in the dynamical evolution of the stellar system, which can be quite
strong for dwarfs on eccentric orbits in an external gravitational
potential~\citep[e.g.,][]{BM2003,Reaetal2006}. In particular, the old-age population of Carina is widely extended and
thus is more likely to be tangentially-biased by tidal effects. Furthermore, \citet{Loketal2010}~considered a scenario
where a dSph could have originated from disky dwarf galaxy whose disk was transformed into a bar-like structure
and tidally stripped by the effects of the Galactic gravitational potential. In this scheme, the velocity anisotropy
profile of the bar-like structure is radially biased in the inner part, while in the outskirts the tidal effects would
dominate so that the profile should decrease toward tangentially biased. This prediction is very good agreement with
the velocity anisotropy profile of Carina found here as well as its velocity structure measured by~\citet{Fabetal2016}
which included the detection of remnant rotation in the dwarf. In addition, using photometric observations
\citet{Batetal2012} found evidence for the presence of tidal tails and isophote twists in Carina, which also provide
observational evidence in favor of tidal interactions. In spite of large uncertainties in our estimated velocity
anisotropy, our results seem to catch a glimpse of kinematical features due to tidal effects from the Galactic
potential.

%%% Sec.7 %%%%%%%%%%%%%%%%%%%%%%%%%%%%%%%%%%%%%%%%%%%%%%%%
\section{Concluding Remarks}

In this work, based on axisymmetric Jeans equations, we constructed non-spherical dynamical models taking
into account multiple components within a stellar system. We assumed that these multiple stellar populations with
different dynamical properties settle in the same gravitational potential provided by a dark halo. We used
a joint likelihood function which is a combination of likelihood functions for each population but assumed
the same dark halo parameters. Applying this dynamical modelling technique to the Carina dSph galaxy
whose stellar populations can be separated by the photometric color magnitude diagram, we found several
interesting results concerning the dark halo and kinematic properties of Carina.

Firstly, in our joint analysis, the halo parameters $b_{\rm halo}$ and $\rho_{0}$ are better constrained than the other
parameters. In comparison with previous works, both parameters in our results have significantly higher values. For the
inner slope of dark matter density, although the uncertainties of this parameter are still large, we have found that
for the Carina dark halo a shallow cusped or cored dark matter profile is preferred. On the other hand,
$(Q,\alpha,\beta,i)$ are very widely distributed in the parameter spaces, and it is difficult to constrain these
parameter distributions, even in joint analysis. We believe that for the joint analysis to work
better and give stronger limits on the dark halo parameters a larger number of observed stars in the kinematic sample
is absolutely essential. Still, the joint analysis enables us to investigate velocity anisotropy profiles
separately for each of the populations.

Secondly, using our fitting results, we have estimated astrophysics factors for dark matter annihilation and decay. We
found that these factors for Carina have the highest values among those of classical dSphs
thereby suggesting that this galaxy is one of the most promising detectable targets among
the classical dwarf galaxies for an indirect search with $\gamma$-ray and X-ray observations. We have also calculated
stellar velocity anisotropy profiles for intermediate- and old-age populations and found that both are radially
anisotropic in the inner region, while the former is approaching isotropy and the latter becomes mildly
tangentially biased in the outer regions. Although these estimates of velocity anisotropy are still subject to large
uncertainties, this feature provides us with the observational kinematic evidence of
tidal effects from the Galactic potential.

So far, the dark matter density profile of the Carina dSph galaxy still has large uncertainties because the dark halo
parameters, especially $Q,\beta_z$~and $i$, are not strongly constrained. To improve these estimates, we require not
only a considerable amount of photometric and spectroscopic data but also proper motions of the member stars within the
galaxy. In particular, the only one way to break the degeneracy between the shapes of the dark halo and the stellar
velocity anisotropy is still to measure three-dimensional velocities of stars. Further observational progress
implementing space and ground-based telescopes will enable us to measure a huge number of stellar kinematic and
metallicity data and, in the more remote future, provide phase space information for stellar systems, thereby
allowing us to set more robust constraints on the dark matter density profile in dSphs.

\section*{Acknowledgements}
We are grateful to the referee for her/his careful reading of our paper and thoughtful comments.
This work was supported in part by the MEXT Grant-in-Aid for Scientific Research on Innovative Areas, No.~16H01090, No.~18H04359 and No.~18J00277 (for
K.H.), by the Polish National Science Centre under grant 2013/10/A/ST9/00023 (for E.L.{\L.}) and by grant AYA2014-56795-P of the Spanish Ministry of economy and Competitiveness~(for M.M.).

% Don't change these lines
\bsp	% typesetting comment
\label{lastpage}
\end{document}